\documentclass[twocolumn,showpacs,aps,prl,amsmath,amssymb,floatfix,%
superscriptaddress]{revtex4}
\usepackage{graphicx}% Include figure files
\usepackage{dcolumn}% Align table columns on decimal point
\usepackage{bm}% bold math
\draft

\begin{document}
\title{Spin Gap in Chains with Hidden Symmetries}
\author{M.N. Kiselev}
\affiliation{Institut f\"ur Theoretische Physik, Universit\"at
W\"urzburg, D-97074 W\"urzburg, Germany}
\author{D.N. Aristov}\altaffiliation[On leave from ]
{Petersburg Nuclear Physics Institute, Gatchina  188300, Russia.}
\affiliation{ Max-Planck-Institut f\"ur Festk\"orperforschung,
Heisenbergstra\ss e 1, 70569 Stuttgart, Germany}
\author{K. Kikoin}
\affiliation{ Ben-Gurion University of the Negev, Beer-Sheva
84105, Israel}

\date{\today}
\begin{abstract}
We investigate the formation of spin gap in one-dimensional models
characterized by the groups with hidden dynamical symmetries. A
family of two-parametric models of isotropic and anisotropic
Spin-Rotator Chains characterized by $SU(2)\times SU(2)$ and
$SO(2)\times SO(2)\times Z_2\times Z_2$ symmetries is introduced
to describe the transition from $SU(2)$ to $SO(4)$
antiferromagnetic Heisenberg chain. The excitation spectrum is
studied with the use of the Jordan-Wigner transformation
generalized for $o_4$ algebra and by means of bosonization
approach. Hidden discrete symmetries associated with invariance
under various particle-hole transformations are discussed. We show
that the spin gap in $SRC$ Hamiltonians is characterized by the
scaling dimension $2$$/$$3$ in contrast to dimension $1$ in
conventional Haldane problem.
\end{abstract}

\maketitle More than 20 years ago Haldane \cite{hal} made a
conjecture that the properties of spin $S$ Heisenberg
antiferromagnetic (AF) chains are different for integer and
half-integer spins. Namely, the excitations in the Heisenberg AF
chains with half-integer spins are gapless, whereas for integer
spins there is a gap in the spectrum (Haldane gap). While the
first part of Haldane conjecture has been proven long time ago
(see \cite{aflieb,lsm}), the second part, although being confirmed
by many numerical \cite{num} and experimental \cite{exp} studies
and tested by some approximate analytical calculations
\cite{lut1}-\cite{ham} remains  a hypothesis. The problem of
$SU(2)$ Heisenberg chains has been attacked by the modern tools as
e.g. bosonization \cite{lut1}-\cite{sch1} (see also book
\cite{gnt}), various numerical methods \cite{num, bot, kit} and
recently proposed fermionization by means of Jordan-Wigner
transformation for higher spins \cite{bo1}. However, the main
focus of interests has been put either on  $S$$=$$1$ chains
characterized by $SU(2)$ symmetry or on $N$-leg ladders described
in terms of dynamic $SO(N)$ groups \cite{poil}. There have also
been made several conjectures concerning spontaneous discrete
symmetry breaking in $S$$=$$1$ chain models associated with e.g.
existence of hidden $Z_2$ and $Z_2$$\times$$Z_2$ symmetries
\cite{kit,ken}. Nevertheless, the general question about the
nature of spin gap is still open.

In this paper we propose yet another approach to the spin gap
problem. It is based on investigation of a family of
two-parametric Hamiltonians described by dynamical groups
\cite{kka}. This family includes conventional two-leg ladder and
several models intermediate between the ladder and the chain. Here
we concentrate on the most instructive example of a
"barbed-wire-like" chain with spins 1/2 in each site coupled by
the ferromagnetic exchange $J_\perp$ within a rung and the AF
interaction $J_\parallel$ along the leg (Fig.1). The model
Hamiltonian  is
\begin{figure}
\includegraphics[width=0.4\textwidth]{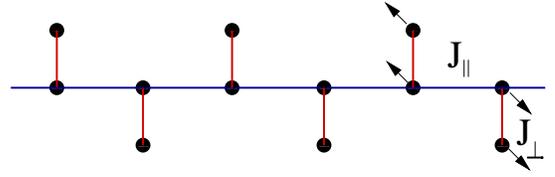}\\
\caption{\label{fig:f1a} Spin Rotator Chain. }
\end{figure}
\begin{equation} H=J_\parallel\sum_i
\vec{s}_{1,i}\vec{s}_{1,i+1}-J_\perp\sum_i\vec{s}_{1i}\vec{s}_{2i}.
\label{h0}
\end{equation}

This model is a natural extension of the $S=1$ chain model to a
case where the states on a given rung form a triplet/singlet pair.
We call the chain shown in Fig. \ref{fig:f1a} the Spin Rotator
Chain ($SRC$) (in contrast to the spin rotor model
\cite{ham,Sach}). Unlike  earlier attempts to construct the
representation of $S=1$ state out of $s=1/2$ ingredients
\cite{lut3,sch1}, we respect in this case the $SO(4)$ symmetry of
spin manifold on each rung \cite{KA}. As a result, {\it the
singlet state  cannot be projected out}. Moreover, it plays
integral part in formation of the spin gap. We show that the
hidden $Z_2$ symmetries in this model are the intrinsic property
of the local $SO(4)$ group of spin rotator  on  the rung, and the
symmetry breaking due to nonlocal (string) effects results in spin
gap formation. These special symmetries distinguish our model from
$N\ge 2$-leg ladder models and $SU(2)$ chains. In particular we
show also that the scaling dimension of a spin gap in SRC differs
from that in 2-leg ladder.

New variables on a rung are introduced to keep track on $S=1$
properties . We define $ \vec{S}_i=\vec{s}_{1,i}+\vec{s}_{2,i},\;
\vec{R}_i=\vec{s}_{1,i}-\vec{s}_{2,i},$ where $\vec{S}_i$ stands
for a triplet $S=1$ ground state and singlet $S=0$ excited state.
The operator $\vec{R}$ describes dynamical triplet/singlet mixing
\cite{kka, KA}. Then
\begin{eqnarray}
H &=& \frac{J_\parallel}{4}\sum_i\left[ \vec{S}_{i}\vec{S}_{i+1}
+\vec{S}_{i}\vec{R}_{i+1}+ (\vec{S}\leftrightarrow \vec{R})\right]
\nonumber\\
&&-\frac{J_\perp}{4}\sum_i\left(\vec{S}_{i}^2-\vec{R}_{i}^2\right),
\label{ham1}
\end{eqnarray}
where the set of operators $\vec{S}_i,\vec{R}_i$ fully defines the
$o_4$ algebra in accordance with the  commutation relations
\begin{eqnarray}
[S_i^\alpha,S_j^\beta]=i\delta_{ij}\varepsilon_{\alpha\beta\gamma}S_i^\gamma,
\;\;
[R_i^\alpha,R_j^\beta]=i\delta_{ij}\varepsilon_{\alpha\beta\gamma}S_i^\gamma,
\nonumber
\end{eqnarray}
\vspace*{-6mm}
\begin{eqnarray}
[R_i^\alpha,S_j^\beta] =
i\delta_{ij}\varepsilon_{\alpha\beta\gamma}R_i^\gamma,\;\;
\label{alg1}
\end{eqnarray}
where $\epsilon_{\alpha\beta\gamma}$ is the totally antisymmetric
Levi-Civita tensor and Casimir constraints on each sites are given
by
\begin{equation}
(\vec{S_i})^2+(\vec{R_i})^2=3,\;\;\;\;\;\;(\vec{S_i}\cdot\vec{R_i})=0.
\label{cas1}
\end{equation}
In order to characterize low-lying excitations in SRC  we propose
a fermionization procedure, which extends Jordan-Wigner (JW)
transformation to $SO(4)$ group, and a bosonization formalism
based on this procedure. Our method incorporates JW transformation
for $S=1$ proposed by Batista and Ortiz (BO) in \cite{bo1}. The BO
representation is however redundant and requires a constraint
overlooked in \cite{bo1}.  The relationships between $SO(4)$ JW
representation and BO representation is discussed below in some
detail.

We begin with a single-rung dimer problem. A two-component fermion
$(a^\dagger b^\dagger)$ basis representing $\vec{S}$-operators is
introduced as follows ($S^\pm=S^x\pm iS^y$)
$$
S^+=a^\dagger +  e^{i\pi a^\dagger a}
b^\dagger,\; S^-=a + b e^{-i\pi a^\dagger a},\;S^z=a^\dagger a
+b^\dagger b -1.
$$
The complementary representation for $\vec{R}$
generators is
$$
R^+=a^\dagger -  e^{i\pi a^\dagger a} b^\dagger,\; R^-=a - b
e^{-i\pi a^\dagger a},\; R^z=a^\dagger a - b^\dagger b.
$$
This representation satisfies commutation relations (\ref{alg1})
for the $SO(4)$ group and preserves Casimir operators
(\ref{cas1}). The advantage of two-fermion formalism in comparison
with two independent JW transformations for each $s=1/2$ is that
the latter requires an additional Majorana fermion to provide
commutation of two spins on the same rung. Two-component spinless
fermions may be combined into one spin fermion, which is most
conveniently done by the definition
\begin{equation}
f_\uparrow=(a-b)/\sqrt{2},\;\;\;\;\;
f^\dagger_\downarrow=(a+b)/\sqrt{2}. \label{ff}
\end{equation}
In order to generalize one-rung representation for a linear chain
of rungs we introduce a "string" operator $K_j$
\begin{equation}
K_j=\exp[i\pi \sum_{k<j,\sigma}n_{\sigma k}]=
\prod_{k<j}(1-2n_{\uparrow k})(1-2n_{\downarrow k}),\label{string}
\end{equation}
($n_\sigma=f^\dagger_\sigma f_\sigma$). As a result of JW
transformation the $SO(4)$ generators acquire the following form
\begin{eqnarray}
S^+_j&=&\sqrt{2}\left(f_{\uparrow j}^\dagger(1-n_{\downarrow
j})K_j+K_j^\dagger f_{\downarrow j}(1-n_{\uparrow j})\right),
\nonumber\\
S_j^-&=&\left(
S_j^+\right)^\dagger,\;\;\;\;\;S^z_j=n_{\uparrow j}-n_{\downarrow
j}, \label{s}\\
R^+_j &=&
\sqrt{2}\left(f_{\uparrow j}^\dagger n_{\downarrow
j}K_j+K_j^\dagger f_{\downarrow j}
n_{\uparrow j}\right), \nonumber\\
R_j^-&=&\left(R_j^+\right)^\dagger,\;\;\;\;\;R^z_j=f_{\uparrow
j}^\dagger f_{\downarrow j}^\dagger +f_{\downarrow j} f_{\uparrow
j}.\label{rpm}
\end{eqnarray}
Part of the representation (\ref{s}) describing $S$$=$$1$
coincides with BO representation. Nevertheless, since $\vec{S}^2$
is no more a conserved quantity, being defined by $
\vec{S}_j^2$$=$$2[$$1$$-$$n_{\uparrow j}$$n_{\downarrow j}$$]$,
the projection of $SO(4)$ group on $S$$=$$1$ representation of
$SU(2)$ group requires an additional Hubbard-like interaction
responsible for the hidden constraint overlooked in BO paper
\cite{bo1}. When the $S$$=$$1$ sector is fixed, three states
$(n_\uparrow,n_\downarrow)$, namely $(1,0)$, $(0,0)$ and $(0,1)$
determine three-fold degenerate triplet state whereas the doubly
occupied state $(1,1)$ stands for a singlet separated from the
ground state by the gap $\Delta$$=$$J_\perp$. The Hamiltonian
(\ref{ham1}) is fermionized by means of purely 1D string operator
$K_j$ (\ref{string}) in contrast to meandering strings proposed
for the theory of 2-leg ladders (see \cite{nun} and references
therein).

The Hamiltonian of anisotropic $XXZ$ SRC model is
$H=H_\parallel+\sum_{i}H_{\perp,i}$, where
\begin{eqnarray}
&&H_\parallel =\frac{J^{x}_\parallel}{8}\sum_i\left[
S^+_{i}S^-_{i+1} +S^+_{i}R^-_{i+1}+ (S\leftrightarrow
R)+h.c.\right]\;\;\;\;\;\;\;\;\; \nonumber\\
&&\;\;\;\;\;\;+\frac{J^{z}_\parallel}{4}\sum_i\left[
S^z_{i}S^z_{i+1}
+S^z_{i}R^z_{i+1}+ (S^z\leftrightarrow R^z)\right]
\label{ham2}\\
&&H_{\perp,i}=\frac{J^{x}_\perp}{8}\left(R^+_i
R^-_i+R^-_iR^+_i\right)+\frac{J^{z}_\perp}{4}(R^z_i)^2-(\vec{R}_i\leftrightarrow
\vec{S}_i). \nonumber
\end{eqnarray}
There exists a set of discrete transformations keeping the
Hamiltonians (\ref{ham1}) and (\ref{ham2}) intact and preserving
commutation relations (\ref{alg1}) and Casimir operators
(\ref{cas1}). In general, these transformations are described by
the matrix of finite rotations characterized by Euler angles
$\theta, \psi, \phi, \varphi$ for the case of $SU(2)\times SU(2)$
or $SO(2)\times SO(2)\times Z_2 \times Z_2$ groups. An example of
such transformation is
\begin{equation}
S^+\to R^+,\; S^-\to R^-,\; S^z\to S^z,\; R^z\to R^z.
\end{equation}
being a $U(1)\times U(1)$ rotation in "$S^x-R^x$" and "$S^y-R^y$"
subspaces. This is in fact a particle-hole flavor transformation
$f_\uparrow\to f^\dagger_\downarrow$, $f_\downarrow\to
f^\dagger_\uparrow$. On the other hand, it corresponds to
replacement $b\to -b$ thus manifesting hidden $Z_2$ symmetry. This
means that an additional gauge factor $\exp(i\theta)$ with
$\theta=\pm \pi$ appears in a fermion operator characterizing
"free ends" of rungs in $SRC$ chain. Other examples are
$(f_\uparrow\to f_\downarrow)$ and $(f_\uparrow\to
f^\dagger_\uparrow, f_\downarrow\to f^\dagger_\downarrow)$. The
latter one corresponds to a particle-hole transformation $(a\to
a^\dagger, b\to b^\dagger)$ in the non-rotated fermion basis.

After JW transformation in $a-b$ basis (\ref{ff})-(\ref{rpm}) the
Hamiltonian (\ref{ham2}) is written as follows
\begin{eqnarray}
H_\parallel
&=&J^{x}_\parallel\sum_i\left( a^\dagger_i
a_{i+1}+a^\dagger_{i+1} a_i \right)\cos(\pi n^b_i) \nonumber\\
 &+& J^{z}_\parallel\sum_i\left(
n^a_i-\frac{1}{2}\right)\left(n^a_{i+1}-\frac{1}{2}\right)
\label{par1}
\end{eqnarray}
and $H_\perp=\sum_i H_{\perp, i}$ with
\begin{equation}
H_{\perp,i}=-\frac{J^{x}_\perp}{2}(a^\dagger_i b_i + b^\dagger_i
a_i)- J^{z}_\perp( n^a_i-\frac{1}{2})(n^b_{i}-\frac{1}{2})
\label{perp2},
\end{equation}
where the shorthand notations $n^a=a^\dagger a$, $n^b=b^\dagger b$
and $\cos(\pi n^b)={\rm Re} \exp(\pm i \pi n^b)=1-2 n^b$ are used.
Below we consider the domain $J_\perp \ll J_\|$ where strongest
deviations from conventional Haldane gap regime \cite{lut1,lut3}
are anticipated. In the limit $J_\perp=0$ our $SRC$ model reduces
to an $s=1/2$ AF chain, the gauge factor $\cos(\pi n^b)=\pm 1$ is
a fictitious random variable which can be eliminated by $S^x\to
-S^x$ and $S^y \to -S^y$ on the corresponding site. This situation
is similar to the so called Mattis disorder \cite{mat} where the
randomness in interaction is removed by proper redefinition  of
spin variables.

The kinematic factor $\sim \cos(\pi n^b_i)$ in $H_\|^x$
(\ref{par1}) can be eliminated by a unitary transformation $\tilde
H$$=$$U^\dagger$$H$$U$ with $U$$=$$\exp$$($$i$$\pi$$\sum_{l,j>l}$$
n^a_j$$n^b_l$$)$. Then $H_\bot^z$ and $H_\|^z$ remain unchanged
and $J^x_\perp$ term acquires the  string form
\begin{equation}
\tilde H^x_{\perp,i}=-\frac12{J^{x}_\perp}\left(a^\dagger_i b_i
e^{-i\pi\sum_{j<i}[a^\dagger_j a_j+b^\dagger_j b_j]}+ h.c.\right).
\label{parper}
\end{equation}
The $s=1/2$ chain is represented in terms of a half-filled band of
fermions. Since interactions (\ref{par1}) and (\ref{perp2}) do not
change the occupation numbers for each color, we expect that
interacting case is also represented by two half-filled bands (see
below). We note that the Hamiltonian $H_\parallel$ in (\ref{ham2})
possesses $U(1)\times U(1)$ symmetry whereas only one local $U(1)$
associated with $b$-fermions exists in (\ref{par1}) due to
non-local character of JW transformation.
\begin{figure}
\includegraphics[width=0.3\textwidth]{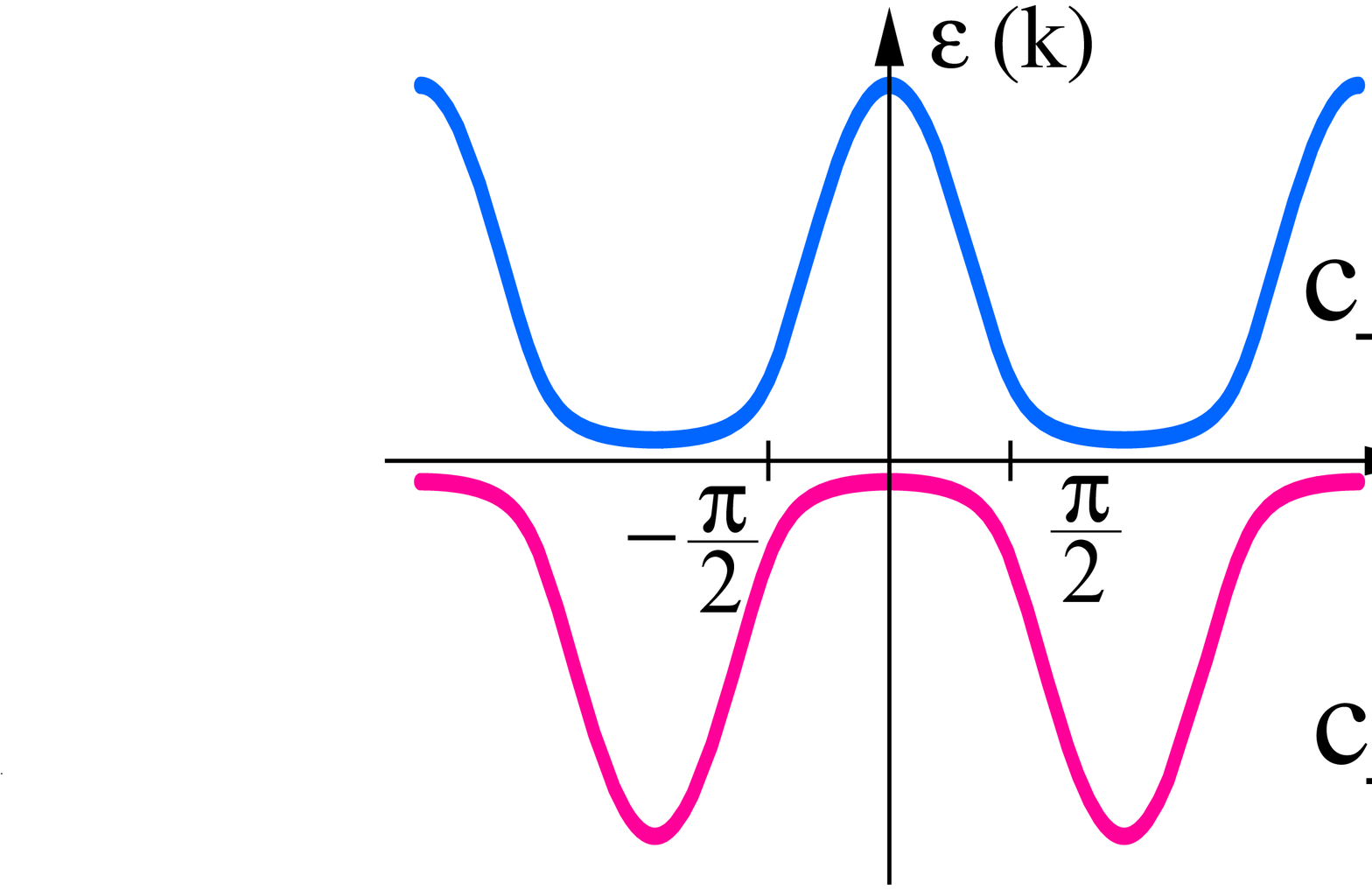}\\
\caption{\label{fig:f2a} Dispersion law for hybridized spin
fermions $c_\pm$.}
\end{figure}

Let us consider the $XY_\parallel-XY_\perp$ model
($J_\parallel^z=J_\perp^z=0$). We spilt the first term in
(\ref{par1}) into the bare hoping and the kinematic term $\sim
J_\parallel^x n^b_i(a^\dagger_i a_{i+1}+h.c)$ playing part of
effective interaction $H_{int}^{XY}$. One gets after
diagonalization of the hopping term
\begin{equation}
H_0=\sum_{p,\lambda=\pm}
\varepsilon_\lambda(p)c^\dagger_{\lambda,p}c_{\lambda,p}
\end{equation}
with $c_+=u_+ a + u_- b,\;\;\; c_-=u_+ b - u_- a$,
\begin{eqnarray}
u_\pm^2(p)&=&\pm
\varepsilon_{\pm}(p)/(\varepsilon_{+}(p)-\varepsilon_{-}(p)),
\label{uv}\\
\varepsilon_{\pm}(p)&=&J_\parallel^x\cos p\pm [ (J_\parallel^x\cos
p)^2+ (J^x_\perp )^2 ]^{1/2}. \label{spectr}
\end{eqnarray}
The chemical potential $\eta$$=$$0$ is pinned in the gap. Thus,
the mixing term fixes global phase difference for $a-b$ - fields.
The remaining symmetry is local $Z_2$$\times$$Z_2$.

We represent $H^{XY}_{int}$ in terms of new variables $c_\pm$ by
expanding the Hamiltonian (\ref{par1}) in the vicinity of two
Fermi points of non-hybridized  system
\begin{equation}
H^{XY}_{int}=\frac{1}{2}\sum_{\{\mu,\nu,\alpha\}=\pm 1, q}
g_{\mu\mu'}^{\nu\nu'}(q) \rho_{\mu\mu',\alpha}(q)
\Lambda_{\nu\nu',\alpha'}(-q) \label{ham3}
\end{equation}
where the
operator $\rho_{\mu\mu'}$ is given by
\begin{equation}
\rho_{\mu\mu',\alpha}(q)=\sum_k c^\dagger_{\alpha,\mu,
k-q/2}c_{\alpha,\mu', k+q/2}. \label{r}
\end{equation}
Its diagonal part is the quasiparticle density. The operator
$\Lambda_{\nu\nu'}$ is defined as
\begin{equation}
\Lambda_{\nu\nu',\alpha}(q)=-\alpha \sum_k k\;\;
c^\dagger_{\alpha,\nu, k-q/2}c_{\alpha,\nu', k+q/2},
\label{j}
\end{equation}
while  its diagonal  part is $\Lambda_{\nu\nu}=
{\rm div} j_{\nu\nu}=-\partial_t \rho_{\nu\nu}$.

In expressions (\ref{r}), (\ref{j}) the index $\alpha=\pm$ stands
for "old" Fermi surface points $k_F^{\pm}=\pm\pi/2$ (we take a
unit lattice spacing), $k$ is measured from $k_F$. Indices $\mu$,
$\mu'$, $\nu$, $\nu'$$=$$\pm$ denote the branch of fermions
$c_\pm$. We used the property of $u_{\pm, \alpha}$$(\pm
\pi/2)$$\approx$$1/\sqrt{2}$. The tensor $g_{\mu\mu'}^{\nu\nu'}$
for these scattering processes has the form
\begin{equation}
g_{\mu\mu'}^{\nu\nu'}=J^x_\parallel(\delta_{\mu\mu'}+\sigma^x_{\mu\mu'})
(\delta_{\nu\nu'}-\sigma^x_{\nu\nu'}). \label{gg}
\end{equation}

We analyze (\ref{ham3}) in terms of $g$-ology approach
\cite{solyom} classifying various terms in
$g_{\mu\mu'}^{\nu\nu'}(q)$ as forward, backward scattering and
Umklapp processes. We see, first, that if $|q|$$\ll$$\pi/2$, and
$g$$\sim$$\pm$$J^x_\parallel$, both diagonal and off-diagonal
matrix elements of $\Lambda_{\nu\nu'}$ vanish in accordance with
Adler's principle \cite{adler}. Thus, the forward scattering
processes leading to small renormalization of the coupling $\sim
(J_\perp^x)^2/J_\parallel^x$ are irrelevant. The backward
scattering processes $($$\pm \pi/2$$\to
$$\mp $$\pi/2$$)$ result in a reduction $J_\parallel$$\to$$
J_\parallel^{1-\gamma}$$J_\perp^\gamma$ of the effective coupling
($0$$<$$\gamma$$<$$1$ is a constant). To get this estimate we cut
logarithmic corrections to the coupling constant by
$\Delta_{min}$$\sim$$ (J^x_\bot)^2$$/$$J_\parallel$ where
$\Delta_{min}$ determines the gap in the density of spin-fermion
states $\varepsilon_\pm$. However there is yet another energy
scale $\Delta$$\sim$$J^x_\perp$ associated with the gap in a
two-point particle-hole correlator with zero total momentum of the
pair. This energy scale is attributed to the gap separating
$S$$=$$0$ excited state on a rung from the triplet state. The
crossover between two energy scales will be discussed elsewhere.
The Hamiltonian (\ref{ham3}) allows also "inter-band" Umklapp
processes determined by the off-diagonal elements of
$\rho_{\mu\mu'}$ and $\Lambda_{\nu\nu'}$ and responsible for
periodicity $Q$$=$$2\pi$. These processes, associated with the
transfer of pair of quasiparticles over the gap do not change the
leading term in (\ref{spectr}).

The above arguments are complemented by the  bosonization
calculations for the strongly asymmetric 2-leg ladder with finite
Fermi velocity $u_b$ in the $b$ subsystem which may be turned to
zero in the end of scaling procedure. The continuum representation
for spin operators $\vec{s}_1, \vec{s}_2$ in (\ref{h0}) reads
\cite{gnt,osa} (we denote ($i$$=$$a(1)$, $b(2)$))
          \begin{eqnarray}
          s_i^\pm(x)&\sim& e^{\pm i\theta_i }(\cos(\pi x)+ \cos(2\phi_i)),
           \nonumber \\
          s_i^z(x)&\sim& \pi^{-1} \partial_x\phi_i  + \cos(\pi x +2\phi_i)
          \label{spinRep}
          \end{eqnarray}
with canonically conjugated variables $\phi_i$ and
$\Pi_i$$=$$\partial_x $$\theta_i$. Keeping only most relevant
terms in the rung interaction, $J^\alpha_\perp$, we arrive at the
conventional equations of Abelian bosonization for the spin
Hamiltonian (\ref{ham1})
          \begin{eqnarray}
          H &=&  \sum_{i=a,b} \int dx\,
        [\frac {\pi u_i K}2 \Pi_i^2 +\frac{u_i}{2\pi K}
         (\partial_x \phi_i) ^2
          \nonumber\\&+&\displaystyle
          J^x_\perp \cos (\theta_a -\theta_b)
          + J^z_\perp\cos2\phi_a \cos2\phi_b]
          \label{ham4}
          \end{eqnarray}
with $K$$=$$1$$/$$2$ and $J_\perp$$\ll$$ u_b$$\ll$$ u_a$$=$$\pi
$$J_\|$$/$$2$ for $J_\|$$ =$$J^x_\|$$ =$$J^z_\|$.

To find the scaling dimension of the gap we start with the case
$J^x_\perp$$=$$0$, $J^z_\perp$$=$$J_\perp$$\neq$$0$. Using the
scaling procedure ($x \to \Lambda x$, $t\to \Lambda t$), one has
$\tilde J_\perp\to J_\perp \Lambda ^{2-\beta}$ where
$\beta$$/$$2$$=$$K$$/$$2$ is a scaling dimension of $\cos2\phi_i$.
The renormalization of $b$-component stops when the renormalized
$J_\perp$ becomes comparable with the lower scale of the energy
$u_b$. The corresponding scale $\Lambda = \xi_b$ defines the first
correlation length $\xi_b =(u_b/J_\perp)^{1/(2-\beta)}$ and the
first energy gap $\Delta_b = u_b \xi_b^{-1} = u_b
(J_\perp/u_b)^{1/(2-\beta)}$. At the second stage of the
renormalization, with frozen $\langle \cos 2\phi_b \rangle \sim
\xi_b ^{-\beta/2}$  the factor $\cos 2\phi_a$ undergoes further
enhancement. The procedure halts when the renormalized amplitude
$\tilde J_\perp$ compares with $u_a$ at $J_\perp \Lambda
^{2-\beta/2}\xi_b ^{-\beta/2} \sim u_a$, which defines a second
correlation length $\xi_a = (\xi_b ^{\beta/2}
u_a/J_\perp)^{1/(2-\beta/2)}$ and the second gap $\Delta_a = u_a
\xi_a^{-1}$. In our particular case $\beta=1$ these formulas
simplify as follows:
\begin{eqnarray}
  \xi_b &=&   (u_b/J_\perp), \quad
\Delta_b = J_\perp \nonumber\\
  \xi_a &=& \xi_b  (u_a /u_b)^{2/3}, \quad
  \Delta_a = J_\perp  (u_a /u_b)^{1/3}.\label{scale}
  \end{eqnarray}
One may decrease $u_b$ in the regime of frozen $\phi_b$ down to
$u_b$$\sim$$J_\perp$. Then
$\Delta_a$$\sim$$J_\|$$($$J_\perp$$/$$J_\|$$)^{2/3}$. Further
decrease of $u_b$ does not change the exponent $2/3$ of the spin
gap fully determined by the scattering on the random potential
$\cos 2\phi_a$ \cite{gia}. The two-stage renormalization procedure
is essential for understanding the SRC model. In the limit
$u_a\sim u_b$, Eq. (\ref{scale}) leads to standard scaling of the
spin gap $\Delta \sim J_\perp$ (see e.g. \cite{ners}).

In the case $J^x_\perp$$\neq$$0$, $J^z_\perp$$=$$0$ the scaling
behavior of the spin gap
$\Delta$$\sim$$J_\|$$($$J^x_\perp$$/$$J_\|$$)^{2/3}$ is determined
by the backward scattering processes of the field $a$ on the
random potential associated with fluctuations of $\cos\theta_a$.

The fully isotropic case, $J^x_\perp$$=$$J^z_\perp$$=$$J_\perp$,
might be expected to yield the same estimate
$\Delta$$\sim$$J_\parallel^{1/3}$$($$J_\perp$$)^{2/3}$. A refined
analysis (see, e.g. \cite{CBV}) including the less relevant terms
in (\ref{ham3}) may correct the gap values, but does not change
this estimate.

To summarize, we introduced a new 1D model intermediate between
the spin $S$$=$$1$ chain and the 2-leg ladder. Our $SRC$ possesses
special hidden $Z_2$ symmetries connected with discrete
transformations in a 6D space of $SO(4)$ group characterizing the
spin rotator. The $SRC$ chain is mapped on the two-component
unconstrained interacting fermions by means of $o_4$ JW
transformation. Two fermion fields are characterized by sharply
different dynamics. One of two fields is frozen at
$k$$\to$$\pm$$\pi/2$ and the scaling dimension $\beta$ of the rung
operator exchange $J_\perp$ is $\beta$$=$$1/2$ instead of
$\beta$$=$$1$ \cite{ners,rdg} in a conventional Haldane problem.
As a result, new scaling "two third" law for the spin gap arises.

We are grateful to N. Andrei, A. Finkel'stein and A. Tsvelik for
useful discussions. This work is supported by SFB-410 project, ISF
grant, A. Einstein Minerva Center and the Transnational Access
program $\#$ RITA-CT-2003-506095 at Weizmann Institute of
Sciences.


\begin{thebibliography}{99}
\bibitem{hal} F.D.M.Haldane, Phys. Rev. Lett. {\bf 50}, 1153 (1983);
Phys. Lett. {\bf 93A}, 464 (1983)
\bibitem{aflieb} I.Affleck and E.H.Lieb, Lett. Math. Phys. {\bf
12}, 57 (1986)
\bibitem{lsm} E.Lieb {\it et al}, Ann. Phys. {\bf 16}, 407
(1961).
\bibitem{num} M.P.Nightingale and H.W.J.Blote, Phys.Rev.{\bf B
33}, 659 (1986), S.R.White and D.A.Huse, Phys. Rev. {\bf B 48},
3844 (1993).
\bibitem{exp}W.J.L. Buyers {\it et al}, Phys. Rev. Lett. {\bf 56}, 371
(1986), J.P.Renard {\it et al}, Europhys. Lett. {\bf 3}, 945
(1987).
\bibitem{lut1} A.Luther and D.J.Scalapino, Phys. Rev. {\bf B 16},
1153 (1977).
\bibitem{lut3} J.Timonen and A.Luther, J.Phys. C {\bf 18}, 1439
(1985).
\bibitem{sch1} H.J.Schulz, Phys.Rev. {\bf B 34}, 6372 (1986).
\bibitem{afl1} I.Affleck {\it et al}, Phys. Rev. Lett. {\bf 59}, 799
(1987).
\bibitem{ham} C.J.Hamer {\it et al}, Phys. Rev {\bf D 19}, 3091 (1979).
\bibitem{gnt} A.O.Gogolin, A.A.Nersesyan and A.M.Tsvelik, {\it
Bo\-so\-ni\-za\-ti\-on and Strongly Correlated Systems}, Cambridge
University Press, 1998.
\bibitem{bot} R.Botet and R.Jullien, Phys. Rev. {\bf B 27}, 613,
(1983).
\bibitem{kit} A.Kitazawa {\it et al}, Phys. Rev. Lett. {\bf 76}, 4038
(1998).
\bibitem{bo1} C.D.Batista and G.Ortiz, Phys. Rev. Lett. 85, 4755
(2001), Advances in Physics, {\bf 53}, 1 (2004).
\bibitem{poil} D.Poilblanc {\it et al}, cond-mat/0402011.
\bibitem{ken} T.Kennedy and H.Tasaki, Phys. Rev. {\bf B 45},304
(1992).
\bibitem{kka} K.Kikoin {\it et al}, cond-mat/0309606.
\bibitem{Sach} S.Sachdev, {\it Quantum Phase
Transitions}, Cambridge University Press, 1999.
\bibitem{KA} K.Kikoin and Y.Avishai, Phys. Rev. Lett. {\bf 86},
2090 (2001).
\bibitem{nun} T.S.Nunner, T.Kopp, Phys.Rev. {\bf B 69}, 104419
(2004).
\bibitem{mat} D.C.Mattis, Phys. Lett. {\bf 56 A}, 421 (1976).
\bibitem{solyom} J.S\'olyom, Adv. Phys. {\bf 28}, 209 (1979).
\bibitem{adler} S.Adler, Phys. Rev. {\bf 137}, B1022 (1965).
\bibitem{osa} I.Affleck and M.Oshikawa, Phys. Rev. {\bf B 60}, 1038
(1999).
\bibitem{gia} T.Giamarchi {\it et al}, Phys. Rev. {\bf B 64}, 245119
(2001).
\bibitem{ners}D.G.Shelton {\it et al}, Phys.Rev. {\bf B 53}, 8521
(1996).
\bibitem{CBV} Shu Chen {\it et al}, Phys. Rev. {\bf B 67}, 054412
(2003).
\bibitem{rdg} E.Dagotto and T.M.Rice, Science {\bf 271}, 618
(1996)
\end{thebibliography}
\end{document}